\documentstyle[preprint,tighten,aps]{revtex}
\begin{document}
\newcommand{\eps}{\varepsilon}
\newcommand{\be}{\begin{equation}}
\newcommand{\ee}{\end{equation}}
\newcommand{\p}{\partial}
\newcommand{\bea}{\begin{eqnarray}}
\newcommand{\eea}{\end{eqnarray}}

\preprint{Commun. Math. Phys. {\bf 166}, 221-235 (1994).}

\title{On the Newtonian Limit of General Relativity}
\author{Simonetta Frittelli\thanks{Fellow of CONICOR}\thanks {current
address:  Physics Department, University of Pittsburgh,
Pittsburgh, PA 15260} \and Oscar Reula\thanks{Member of CONICET}}
\maketitle
\begin{center}
FaMAF, Laprida 854, 5000 C\'ordoba,  Argentina
\end{center}
\vspace{5mm}

\begin{abstract}

We establish rigorous results about the Newtonian limit of general
relativity by applying to it the theory of different time scales for
nonlinear partial differential equations as developed in [4,1,8].
Roughly speaking we obtain a priori estimates for solutions to Einstein
equations, an intermediate, but fundamental, step to show that given a
Newtonian solution there exist continuous one-parameter families of
solutions to the full Einstein's equations --the parameter being the
inverse of the speed of light-- which for a finite amount of time are
close to the Newtonian solution.  These one-parameter families are
chosen via an {\sl initialization procedure} applied to the initial
data for the general relativistic solutions.  This procedure allows one
to choose the initial data in such a way as to obtain a relativistic
solution close to the Newtonian solution in any a priori given Sobolev
norm. In some intuitive sense these relativistic solutions, by being
close to the Newtonian one, have little extra radiation content
(although, actually, this should be so only in the case of the
characteristic initial data formulation along future directed light
cones).

Our results are local, in the sense that they do not include the
treatment of asymptotic regions; global results are admittedly very
important --in particular they would say how differentiable the
solutions are with respect to the parameter--, but their treatment
would involve the handling of tools even more technical than the ones
used here.  On the other hand, this local theory is all what is needed
for most problems of practical numerical computation.

\end{abstract}
\newpage

%%%%%%%%%%%%%%%%%%%%%%%%%%%%

\section{Introduction}

As it has been suggested through the extended history of the treatment
of the newtonian limit of relativity, slow-motion corrections should be
naturally obtained by suitable approximation schemes starting from the
newtonian gravitation theory.  These schemes assume, for their
validity, that there is a sufficient number of sufficiently smooth
one-parameter families of solutions to the full Einstein equations
(from now on referred to alternatively as {\sl the relativistic equations}),
the parameter being the inverse of the speed of light, and smoothness
of these solutions with respect to this parameter being valid even at
its zero value.  Their smoothness is required so that the different
order corrections can be interpreted as Taylor series coefficients of
these one-parameter families. Their existence in sufficient number is
required so that there is at least one for each newtonian solution,
that is, so that we can describe every physical situation that can
occur.  Our purpose is to show some rigorous results about this matter,
that is to show the existence of a sufficiently large number of these
smooth one-parameter families of solutions.

The general theory of gravity coupled with matter sources obeying their
own symmetric hyperbolic equations of motion is a problem of at least
two different time scales, one having as characteristic speed the speed
of light, the other --much slower-- the speed of sound of the matter.
One applies newtonian and post-newtonian approximations when one is
interested in solutions where {\sl things happening at the fastest time
scales are small and unimportant for the bulk motion of the sources.}
We will call them slow-motion solutions.  That is the case for instance
of the solar system; as a first approximation one is not interested in
the details of the gravitational radiation, nor is this radiation
important for knowing the motion of the planets in that approximation;
we claim we can describe it --locally-- with a slow solution.  These
slow solutions are the ones forming the one-parameter families referred
to above.

In recent years a complete theory has been developed for treating
nonlinear partial differential equations with different time scales
[4,1,8], which answers our questions in that general setting; not only
does this theory tell us about the existence of these slow-motion
solutions, but it also gives us a recipe called {\sl initialization} on
how to choose initial data for them. Even more, that recipe also allows
us to choose data whose evolution is arbitrarily close to the
corresponding solution to the limiting system, that is the system of
equations obtained when the largest speed is set to infinity.  Our
approach to the Newtonian limit will then be to set the relativistic
equations in a form suited for the application of the different
time-scales theory above mentioned, apply it and see what kind of
initialization procedure (further constraints on the initial data) we
obtain.  Solutions of this extra constrained system are the slow-motion
solutions we are seeking.

The plan of the paper is as follows:

In the second section we present an introduction to the theory of
quasilinear symmetric hyperbolic equations with different time scales.
We briefly discuss what the basic principles of that theory are, what
the requirements on the system of equations are for it to work, and
finally what the results are.

In the third section we present a new formulation (essentially a
redefinition of variables) of Einstein equations as a first order
symmetric hyperbolic system for arbitrary --but given-- lapse and shift
tensors.  Here we assume the matter fields and their interactions with
gravity to be such that the whole set of equations (including the
matter fields) is block-diagonal symmetric hyperbolic. For instance,
the equations for the matter fields may themselves be symmetric
hyperbolic and depend only on the metric and its connection (and not on
the curvature tensor).  The example we always have in mind as a matter
source is that of a perfect fluid.

In the fourth section we make use of this new formulation, i.e. of the
freedom of choosing arbitrarily the lapse and shift, to pick a
particular gauge in which our system satisfies the requirements of the
different time-scale theory.  This forces us to introduce elliptic
equations on each time slice for the lapse and shift, and therefore to
treat now a mixed hyperbolic-elliptic system of nonlinear equations.

In the fifth section we discuss the initialization procedure, that is
the selection of initial data that give rise to time-regular
solutions.

In the sixth section we summarize the results obtained and discuss how
they might be ~embedded in particular ~settings to yield actual
theorems. We also give our expectation for the local problem and for
the asymptotically flat (global) case.

\section{The Theory of PDE with Different Time Scales}

There follows a short review of the main ideas behind this theory; the
complete and detailed version can be found in [4,1,8]. This treatment
is local in the sense that we take the region for the time evolution of
the problem to be a compact spacelike region $S$ cross the positive
time ($S \times R^+$).  At the time-like boundary of this region, ($\p
S \times R^+$) --which we take to be noncharacteristic--  we assume
there exist suitable boundary conditions which guaranty the uniqueness
of the solutions for given initial data. This is so for linear
equations without constraints, but the non-linear constrained case
--which includes general relativity-- is far from being complete,
although results for nonlinear equations and linear systems with
constraints are available at the present time.

The key step to prove uniqueness and existence of solutions to
hyperbolic systems is to establish an {\sl a priori inequality} which
bounds a certain norm of the assumed solution at some latter time by a
multiple of the norm of the initial data.  This inequality is called
the {\sl energy estimate} because for physically relevant linear
systems the weakest  norm of this type that can be used to obtain this
inequality, and so to assert existence, is just the square root of the
energy.  This a priori energy estimate is:

Given a solution $u^k_0$ which for every $t \in [0,T]$ is in $H^m(S)$,
there exists a constant $C$ such that given any other solution $u^k \in
H^m(S)$ sufficiently close to the first one we have~\footnote{Here we
consider first order systems, and so the initial data is just
$u(0)$.}:

\be
|| u(t) ||_{H^m(S)} \leq C || u(0) ||_{H^m(S)}, \;\; \forall \; t \in
[0,T],
\label{eq:energy}
\ee

where $H^m(S)$ is a generalization of the usual Sobolev spaces of order
$m$, which not only include space derivatives up to order $m$, but also
time derivatives and crossed time-space derivatives up to the same
order.  In spite of that, the norm at $t=0$ is bounded by the usual
Sobolev norm of order $m$ if one uses the evolution equations to trade
all time derivatives appearing there for space derivatives. The bound using
just the initial data --and not their time derivatives-- is the one
that is important in establishing existence of solutions; we shall come
to this again when analyzing systems with different time scales.  The
minimal value of $m$ for which the energy estimate is valid for a
generic (non-linear) symmetric hyperbolic system --a concept we
introduce bellow-- is the smallest integer larger than $n/2 + 2$, where
$n = dim\; S$, for this guarantees that we have pointwise bounds on
$u$, a sufficient step in the non-linear case to obtain the
inequality.

A sufficient condition for a system of evolution equations to obey the
above inequality is that it be symmetric {\sl at a solution $u^k$},
that is, a system that can be  written in  the following form:

\be
A^0{}_{ij}(u^k) \frac{\p}{\p t} u^j = A^a{}_{ij}(u^k)
\nabla_a u^j
+ B_{i}(u^k),
\label{eq:s-h}
\ee

where the matrix $A^0_{ij}$ is symmetric and positive definite and the
vector matrix $A^a{}_{ij}(u^k)$ is symmetric. The above matrices and
$B_i$ are supposed to be smooth functionals of the vector $u^k$. The
connection $\nabla_a$ is some arbitrary connection on $S$.~\footnote{We
are using Penrose's abstract index notation, by which latin
supra-indices denote vectorial type entries of a tensor and sub-indices
co-vectorial ones.}

If we scale some of the components of $A^a{}_{ij}$ with a factor
$1/\eps$, as it happens in systems with different time scales, where
$1/\varepsilon$ is the largest speed of the system, much larger than
the others, then the constant appearing in the energy estimate,
equation (\ref{eq:energy}), will in general appear also scaled with a
$1/\varepsilon$ factor and in the limit $\eps \to 0$ as this factor
goes to zero one losses the estimate.  This is not always the case, and
as shown in [4,1,8] the dependence of $C$ on $\varepsilon$ is regular
if~\footnote{Actually there exist more general and so more
sophisticated conditions; we do not include them here because they are
not needed for our purposes.} both the following conditions are met:

$1.)$ The matrices $A^a{}_{ij}$ appearing in the symmetric hyperbolic
system have the following form:

\be
 A^a{}_{ij}(u^k,\eps)
= \frac{1}{\eps}A^a_0{}_{ij} + A^a_1{}_{ij}(u^k,\eps),
\ee

with $A^a_0{}_{ij}$ being constant matrices, and $A^a_1{}_{ij}$ regular
in $\eps$. The reason for this is simply that in this case the singular
terms go away on integration by parts.

$2.)$ The matrix $A^0{}_{ij}$, and the vector $B_i$ are regular in
$\eps$.

There is another way to loose the estimate, namely if the norm at $t=0$
--i.e. on the initial data-- blows up in the limit (here we refer to
the fact that in trading time derivatives of the initial data for their
space derivatives one again encounters the singular part of
$A^a{}_{ij}$). The only way to avoid this singular behavior is to
choose very special initial data, that is to constrain the initial data
by imposing on them some (elliptic) differential equations that
guarantee the good behavior of the bound on the limit.  This process of
selecting particular initial data, and so the so called slow solutions,
is called {\sl initialization}.  To summarize, the above theory tells
us that if the symmetric hyperbolic system satisfies conditions $1.)$
and $2.)$ above then there are certain solutions coming from
initialized data which depend smoothly on the parameter $\eps$, in the
sense that we get the a priori estimate $(1)$.

In the next section we define appropriate variables for general
relativity in such a way that the resulting matrix $A^a{}_{ij}$
satisfies condition $1.)$.  ~Unfortunately condition $2.)$ is in
general not satisfied, for the resulting vector $B_i$ contain singular
(w.r.t $\eps$) terms. These terms can not be completely eliminated with
gauge conditions, but their ~undesirable effects on the energy
estimates actually can. In section  {\bf 4} we show that all singular
terms on the integral of $u_j B_i(u_k) \delta^{ji}$, and similar terms
appearing in higher order energy expressions, can be ~annihilated by a
particular gauge choice. This gauge fixing is given by elliptic
equations that must be solved in each time slice, but since these
equations depend on the dynamical variables we must then treat a
coupled hyperbolic-elliptic system of equations. We give arguments
showing that, nevertheless, the usual estimates can be obtained.

%%%%%%%%%%%%%%%%%%%%%%%%%%%%

\section{The Symmetric Hyperbolic System}

%%%%%%%%%%%%%%%%%%%%%%%%%%%%

In this section, as a first step to treating Einstein equations as a
system with different time scales, we cast them as a first order
symmetric hyperbolic system. The system we present here has this
remarkable property for arbitrarily given lapse and shift variables,
and was found in collaboration with R. Geroch. Further details of that
work will be published elsewhere.  We shall make use of this property
when studying the newtonian limit, for this requires the choice of a
specific gauge, which only arises as a posteriori consequence of having
the system in a symmetric hyperbolic form. We remark that we were not
able to obtain a regular limit in the harmonic gauge; the same problem
is already present in the corresponding limit for the electromagnetic
field (in the Lorentz gauge) and has to do with the coupling to the
fluid.

%%%%%%%%%%%%%%%%

\subsection{Variables and Equations}

%%%%%%%%%%%%%%%%

We consider now the dynamical problem of general relativity, namely,
the temporal evolution of a 3-dimensional spatial metric. As usually,
we take a $(3+1)$ decomposition of spacetime, so we take it to be
foliated by spacelike surfaces $\Sigma_t$, which are the
level surfaces of a function $t$.  The normal $n_a$ to these surfaces
is $\bar Ndt$ for some function $\bar N$. The spacetime metric $g_{ab}$
induces a metric $\bar q_{ab}$ on the spatial slices: $\bar q_{ab} =
g_{ab} + n_a n_b$.  To this variable corresponds a canonically
conjugate one, its momentum: $ \bar \pi^{ab} = \frac {\sqrt{\bar
q}}{\eps} \,(\bar q^{ab} \bar K^{c}{}_{c} - \bar K^{ab}) $, where $\bar
K^{ab} \equiv \frac12 \hbox{\pounds}_{n^c}\bar q^{ab}$ is the extrinsic
curvature of the surfaces  and $\sqrt{\bar q}$ is the square root of
the determinant of the 3-metric~\footnote{The square root of the
determinant of the metric is just a shorthand for the volume element of
that metric}.

In the variables ($\bar q^{ab},\bar \pi^{ab}$) Einstein equations split
into a set of evolution equations and a set of constraints on the
initial data (see for instance [9] ~\footnote{The definition of
momentum given in this reference has a difference with the momentum
assumed here, but, having this in mind, the set of the equations we
present can be obtained from the equations given in this reference by a
straightforward calculation.}):

\be
\dot{\bar q}^{ab}= -\frac{\bar N}{\sqrt{\bar q}} \,
                    (2\bar \pi^{ab}-\bar q^{ab}\bar \pi^c_c)
                   - 2\bar D^{(a}\bar N^{b)}
\ee

\bea
\eps^2 \dot{\bar \pi^{ab}} & = &
                                 -\sqrt{\bar q} \bar N \,
                                  \{
                                    \bar R^{ab}-\frac12 \bar R\bar q^{ab}
                                    + \frac{1}{\bar  N}
                                     (
                                      \bar q^{ab}\bar D^c\bar D_c \bar N
                                      -\bar D^a\bar D^b \bar N
                                                              )
                                                               \} \nonumber \\
                           &   & +\frac{\bar N\eps^2}{\sqrt{\bar q}} \,
                                  \{
                                    \bar \pi^{ab}\bar \pi^c_c
                                    -2\bar \pi^a_c \bar \pi^{cb}
                                    +\frac12\bar q^{ab}
                                     (
                                      \bar \pi^{cd}
                                      \bar \pi_{cd}
                                      -\frac12\bar \pi^c_c
                                              \bar \pi^d_d
                                                          )
                                                           \} \nonumber \\
                           &   & + 2 S^{ab}\eps^4
                                 + \eps^2 \bar N^c\bar D_c\bar \pi^{ab}
                                 - 2 \eps^2 \bar \pi^{c(a}\bar D_c\bar N^{b)}
\eea

\be
                                                   \bar R
       - \frac{\eps^2}{\bar q}(\bar \pi^{cd}\bar \pi_{cd}
                      -\frac12 \bar \pi^c_c \bar \pi^d_d)
                                            - 4\eps^2\rho = 0
\ee

\be
                                - 2 \bar D_c\bar \pi^{ca} = 4 \eps^2 J^a
\ee

Here, $\bar N^a$ is the shift vector; $(\dot{\;}) \equiv
\hbox{\pounds}_{t^a}$ ($t^a = \frac{\bar N}{\eps} n^a + \bar N^a$) and
$\bar D_a$ is the derivative operator on the slices associated with
$\bar q_{ab}$. The parameter $\eps$ is the inverse of the speed of
light, which here will be taken to be the fastest speed when the above
system is considered a system with different time scales. The way in
which this parameter appears on the equations, at this level, is
determined by dimensional considerations, see the appendix for the
rules by which we assign dimensions to the different tensor
quantities.

Note that in these equations, second derivatives of the metric are
involved, for they appear in $\bar R^{ab}$.

If equations (6), and (7), the so called constraint equations, are
satisfied at any given instant of time --i.e. by the initial data--
then equations (4) and (5) imply they are satisfied at all times.

The tensor fields $\rho$, $J^a$, and $S^{ab}$ that appear in the
equations are the different projections of the energy-momentum tensor
of the matter fields on $\Sigma_t$.  We are assuming in what follows
that the matter fields obey by themselves symmetric hyperbolic
equations which are coupled to the metric in such a way that after we
cast Einstein equations in a symmetric hyperbolic form, then the
whole system would be symmetric hyperbolic.  We therefore  ignore the
matter equations.

%%%%%%%%%%%%%%%%

\subsection{Conformal Transformation and Lapse--Shift Scaling}

%%%%%%%%%%%%%%%%

Let $ q_{ab} \equiv \bar N^2 \bar q_{ab} $ and therefore $ q^{ab}
\equiv \bar N^{-2} \bar q^{ab} $.  This choice of conformal factor
leads us to the following expression of the Ricci tensor of the
conformal metric:

\be
     R_{ab} = \bar{R}_{ab}
              - \frac{1}{\bar N}
                 (
                  q_{ab}\bar D^c\bar D_c \bar N+\bar D_a\bar D_b \bar N
                                                                       )
              + \frac{2}{\bar N^2}\bar D_a \bar N\bar D_b \bar N
\ee

Thus, in the evolution equation for $\bar \pi^{ab}$ second derivatives
of the lapse function appear only in the form of the laplacian $ q^{ab}
\bar D_a \bar D_b $.  This is highly convenient since, as we shall
shortly see, this term can be eliminated in favor of the mass density.

Let

\be
     \bar N \equiv N \sqrt{\bar q}
            \equiv \frac{\sqrt{\bar q}}{1-4\eps^2U},
\ee

where $U$ is an arbitrary function. This choice will help us to get rid
of some combinations of second derivatives of the metric that hamper
the system to become symmetric hyperbolic.  The choice $U \equiv 0$
would correspond to the temporal harmonic gauge, see for instance [5].
Here we don't restrict ourselves to this gauge, for $U$ remains
arbitrary.

Nevertheless notice that, up to first order in $\eps$, we do have the
temporal harmonic gauge.  The $2^{nd}$ order correction to it, $U$,
will be latter identified with the {\it newtonian potential}.  This is
in agreement with Nester and K\"{u}nzle [5].

Second derivatives of the lapse will then be proportional to second
derivatives of the newtonian potential plus other terms involving
second derivatives of the metric, which will be arranged by means of a
redefinition of variables in the next section. In this way, there will
be no second order derivatives of the newtonian potential other than
the laplacian.

The same happens with the curvature scalar ~\footnote{We mention this
in order to have a complete description, although this form of the
scalar constraint is not relevant to the actual computations leading to
the final form (equation 16).}, so the scalar constraint is:

\be
                                                     \bar N^2 R
          + \frac{4}{\bar N} \bar q^{cd}\bar D_c\bar D_d \bar N
   - \frac{2}{\bar N^2}\bar q^{cd}\bar D_c\bar N\bar D_d \bar N
                                                 - 4\eps^2 \rho
                        - \frac{\eps^2}{\bar q}
                           (
                             \bar \pi^{cd} \bar \pi_{cd}
                            - \frac12 \bar \pi^c_c\bar \pi^d_d
                                                              )
                                                                 = 0
\ee

So far, the lapse has been redefined and the conformal factor chosen so
that a new variable $U$ appears conveniently for latter purposes.
There are no restrictions yet on this variable, which doesn't need to
be fixed in order to set a well posed initial value formulation, i.e.
to get a symmetric hyperbolic system.  Restrictions only will come by
means of a gauge fixing procedure to obtain a regular newtonian limit,
thus justifying the association of $U$ with the idea of a {\it
newtonian potential}.

For similar purposes we also re-scale the shift,

\be
         {\bar N}^a \equiv {\eps^2} N^a
\ee

%%%%%%%%%%%%%%%%%%%%%%%%%%%%%%

\subsection{New Variables}

To get the symmetric hyperbolic system we define the following
variables:

\be
     r^{ab}{}_c \equiv
                       \frac{1}{2\eps^3}
                        (
                         \p_cq^{ab}-\frac12 q^{ab}q_{ed}\p_cq^{ed}
                                                                  )
                \equiv
                       \frac{1}{2\eps^3 \sqrt q}\p_c
                                                    (
                                                     \sqrt q q^{ab}
                                                                   )
\ee

\be
     p^{ab} \equiv
                   \frac{1}{\eps^2}\bar \pi^{ab}
\ee

Here, $\p_c$ is the derivative operator associated to a flat $e^{ab}$.
For the purpose of studying the Newtonian limit we introduce here
explicitly the inverse of the speed of light ($\eps$) on the definition
of the variables and formulae. This is of no relevance for obtaining
the symmetric hyperbolic system.

It is expected that $\sqrt q q^{ab}$ ~\footnote{Notice that this metric
density agrees with the dynamical variables of M. Lottermoser in (7).}
will differ from the above flat metric in order $\eps^3$. We shall
assume, and in fact then assert, that $\sqrt q q^{ab} = \sqrt e e^{ab}
+ \eps^3 h^{ab}$, with $h^{ab}$ $\eps$-smooth\footnote{this will
follow ~directly once we show that $r^{ab}{}_c$ is pointwise bounded.}.
Therefore, derivatives
of the metric density will be of order $\eps^3$. One can check that
this is so, for instance, in Schwarzschild.

We make use of the constraints to rearrange terms in the evolution
equations.  With appropriate factors, we add the scalar constraint to
the equation for $\bar \pi^{ab}$ to eliminate the $\bar R$ term, and we
add the vector constraint to the equation for the new variable
$r^{ab}{}_c$. The new terms added in this way are necessary to
symmetrize the system $(p^{ab}, r^{ab}{}_c)$. This, of course, involves
the appearance of extra source terms in the evolution equations:

\bea
\label{eq:p.evol}
                 \dot p^{ab} & = & - q^{\frac34} N^{\frac12} \frac1{\eps}
                                      (
                                       q^{cd}\p_dr^{ab}{}_c
                                       - 2q^{c(a}\p_cr^{b)d}{}_d
                                                                )
                                        - 2q^{\frac12}\frac1{\eps^2} q^{ab}
                                      (
                                       \Delta U - \rho
                                                      )
                                                                   \nonumber \\
                             &   & + 2 S^{ab}
                                   + \eps^2  N^c\p_cp^{ab}
                                   + \eps F^{ab}
                                                (
                                                 \eps, r^{de}{}_c, p^{de},
                                                 \p_c U , \p_c N^d
                                                                  )
\eea

\bea
\label{eq:r.evol}
            \dot r^{ab}{}_c & = & -\frac{N^{\frac12}}{q^{\frac14}}\frac1{\eps}
                                    (
                                     \p_cp^{ab} -2\delta^{(a}_c\p_dp^{b)d}
                                                                          )
                                                                \nonumber \\
                            &   & + \frac1{\eps}
                                     \{
                                       q^{ab} \p_c \p_d N^d
                                       - q^{d(b} \p_c \p_d N^{a)}
                                                                 \}
                                                                 \nonumber \\
                            &   & + \frac{4 N^{\frac12}}{q^{\frac14}}
                                     \frac1{\eps} \delta^{(a}_cJ^{b)}
                                  + \frac2{\eps} q^{ab} \p_c (N\dot U)
                                  + \eps^2 N^d \p_d r^{ab}{}_c
                                                                 \nonumber \\
                            &   & - 2 \eps N q^{ab} N^d \p_c \p_d U
                                                                 \nonumber \\
                             &   & + \eps F^{ab}{}_c
                                                   (
                                                    \eps, r^{ab}{}_c,
                                                    p^{ab}, \p_c U , \dot{U},
                                                    \p_cN^a
                                                           )      .
\eea

Here $\Delta \equiv \bar q^{ab} \p_a \p_b$, $F^{ab}= F^{ab}(\eps,
r^{ab}{}_c, p^{ab}, \p_c U , \p_cN^a )$, \\
$F^{ab}{}_c=F^{ab}{}_c(\eps, r^{ab}{}_c, p^{ab}, \p_c U , \dot
U,\p_cN^a)$, and all other $F$'s appearing in the equations from now on
are smooth pointwise functions of all their arguments.

Of course for the solutions to this system to be solutions of Einstein
equations one needs to impose on the initial data the usual
constraints, which in these variables take the form:

\be
                      \Delta U
 -\frac12 \eps \p_c r^{cd}{}_d = \rho
                                 + \eps^2 F(\eps, r^{ab}{}_c, p^{ab}, \p_c U)
\ee

\be
              - 2 \p_c p^{ca} = 4 J^a
                                + \eps^2 F^a(\eps, r^{ab}{}_c, p^{ab}, \p_c U),
\ee

and the extra one:

\be
                  r^{ab}{}_c =  \frac1{2\eps^3 \sqrt q} \p_c (\sqrt q q^{ab}),
\ee

which if satisfied initially then, --modulo boundary conditions--, hold
at all times. To see this take time derivatives of the constraints and
using the above equations get a symmetric hyperbolic system.  Thus, if
the boundary conditions for the original system, eqn's (4, 14, 15, 16,
17, 18), are ~such as to ensure uniqueness of solutions for this
derived system, then the constraints would remain zero for all times
and the original system is equivalent to Einstein's equations, (eqn's
(4,5,6,7)). It is interesting, and perhaps disturbing, that the
constraints propagate at a different speed than (but numerically
proportional to) light.

We claim that the above system is symmetric hyperbolic, for {\bf any
given} choice of lapse and shift $(U, \; N^a)$.  To see this one can
compute the resulting matrices $A^0{}_{ij}, A^a{}_{ij}$, and check
their symmetry. We faind it convenient to split the matrices into
pieces using compounded subindices. Thus we define, for instance,
$A^a_{pr}$ that part of the matrix $A^a_{ij}$ that acts on $p^{ab}$ and
has image in the space of the $r^{ab}{}_c$'s. We take as the matrix
$A^0_{ij}$, one having only the following nonzero components:
$A^0_{qq} = \delta^p_a \delta^q_b$, (we reserve the first letters of
the ~alphabet for contraction with vectors in the domain and the last
ones for the image), $A^0_{pp} = q^{-1} \delta^p_a \delta^q_b$,
$A^0_{rr} = \delta^p_a \delta^q_b \delta^c_r$.  The only nonzero
components of the resulting $A^l_{ij}$ vector valued matrix are then:

$A^l_{pr} = - \frac{q^{\frac34} N^{\frac12}}{\eps} (\delta^c_l
\delta^p_a \delta^q_b - 2 \delta^c_{(a} \delta^p_{b)} \delta^l_q)$, and

$A^l_{rp} = - \frac{q^{\frac34} N^{\frac12}}{\eps} (\delta^l_r
\delta^p_a \delta^q_b - 2 \delta^l_{(a} \delta^p_{b)} \delta^q_r)$.

{}From these formulae by contraction with two arbitrary vectors, it is
easy to see the symmetry of $A^l_{ij}$.  Alternatively, one can compute
the time derivative of the first norm appearing in the energy estimate
(~\ref{eq:energy}), i.e. the energy, and see that after integration by
parts it only depends on the dynamical fields $(\sqrt q
q^{ab},\;p^{ab},\; r^{ab}{}_c)$, and not on their  derivatives. The
expression for this energy is:

\be
    E(t) = \frac12 \int_{\Sigma_t}
                    \left\{
                           r^{ab}{}_c r_{ab}{}^c
                           + q^{-1} p^{ab}p_{ab}
                                                \right\}
                                                        d\Sigma,
\ee

where we have raised and lowered indices using the conformal metric,
$q_{ab}$.

Its time derivative is:

\be
    \dot E(t) =  - \int_{\Sigma_t}
                    \left\{
                           \dot r^{ab}{}_c r_{ab}{}^c
                           + q^{-1} \dot{p}^{ab}p_{ab}
                                                     \right\}
                                                             d\Sigma
                 + \;\;algebraic \;\;terms \;\; in \;\; the \;\; fields.
\ee

Using equations (\ref{eq:p.evol}) (\ref{eq:r.evol}), and integrating by
parts, we obtain:

\bea
     \dot{E}(t) & = &  - \int_{\Sigma_t}
                          \left\{
                                 \frac1{\eps} r_{ab}{}^c
                                  (
                                   q^{ab} \p_c \p_d N^d
                                   - q^{d(b} \p_c \p_dN^{a)}
                                                            )
                                                             \right.
                                                             \nonumber \\
                &   & + \frac{4N^{\frac 12}}{q^{\frac 14}}
                        \frac1{\eps}r_{ab}{}^aJ^b
                      - \frac2{\eps} r_{ab}{}^c q^{ab} \p_c (N\dot U)
                                                              \nonumber \\
                &   & - 2 \eps N r_{ab}{}^c q^{ab} N^d \p_c \p_d U
                                                              \nonumber \\
                &   & \left.
                       - 2 q^{-\frac12}\frac1{\eps^2} p^{ab}q_{ab}
                           (
                            \Delta U - \rho
                                          )
                           \right\}
                                   \;\; d\Sigma
                                                              \nonumber \\
                &   & + \;\; \eps \;\; regular \;\; algebraic
                        \;\; terms \;\; in \;\; the \;\; fields
                        \;\; and \;\; in
                                                              \nonumber \\
                &   &   \;\;\; first \;\; derivatives \;\; of
                        \;\; U \;\; and \;\; N^a.
\eea

We have  explicitly written the terms which contain second derivatives
of the shift and of the  potential, or are singular in $\eps$, for
latter purposes, although they have no relevance in this section
because here we are still assuming the lapse and shift are freely
given, and we are not yet taking the limit $\eps \to 0$.

This symmetric hyperbolic system, with sources having symmetric
hyperbolic evolution equations (such as a perfect fluid), is sufficient
to assert the existence of solutions to the relativistic equations for
any given lapse and shift, --i.e. for any gauge--, and any given value
of the parameter $\eps$, as small as desired but different from zero.
In what follows we are not going to be interested in existence,
since it can be derived from this system or others systems in the literature.
We are going to be mainly concerned with the $\eps$-smoothness of these
solutions for which we assume existence.

%%%%%%%%%%%%%%%%%%%%%%%%%%%%%%

\section{The Newtonian Limit}

%%%%%%%%%%%%%%%%%%%%%%%%%%%%%%

The singular behavior in $\eps$ of the time derivative of the energy
prevents us from obtaining an energy estimate, equation
(\ref{eq:energy}), with constant independent of $\eps$, and therefore,
as discussed in the second section, we do not control the behavior of
the solutions in the limit $\eps \to 0$. From the viewpoint of the
general theory described on the second section the problem arises
because in our case the $B_i(u^k)$ on equation $(2)$ are in fact
singular with respect to $\eps$. These singular terms can not be
eliminated from equation $(2)$ by any gauge condition, but as we shall
see the singular terms that they generate on the energy estimates can
indeed be eliminated by choosing a convenient gauge.

To see this we do further integration by parts on the expression for
the time derivative of the energy to obtain:

\bea
     \dot E(t) & = &  - \int_{\Sigma_t}
                         \left\{
                                \frac1{\eps} r_{ab}{}^a
                                 [
                                  \frac{4N^{\frac12}}{q^{\frac14}} J^b
                                  - q^{cd} \p_c \p_d N^{b}
                                                          ]
                                                           \right.
                                                          \nonumber \\
               &   & - \frac1{\eps} r_{ab}{}^c q^{ab} \p_c
                        (
                         2N\dot U + \p_d N^d
                                            )
                                                            \nonumber\\
               &   & \left.
                     - 2 \eps N r^{ab}{}_c q_{ab} N^c \Delta U
                     - 2 q^{-\frac12} \frac1{\eps^2}
                         p^{ab}q_{ab}(\Delta U - \rho)
                        \right\}
                                                \;\; d\Sigma
                                                            \nonumber \\
               &   & + \;\; \eps \;\; regular \;\; algebraic
                       \;\; terms \;\; in \;\; the \;\; fields
                       \;\; and \;\; in
                                                             \nonumber \\
               &   &   \;\;\; first \;\; derivatives \;\; of
                       \;\; U \;\; and \;\; N^a.
\eea

We now use the scalar constraint equation to get rid of the laplacian
of $U$, and get:

\bea
     \dot E(t) & = &  - \int_{\Sigma_t}
                         \left\{
                                \frac1{\eps} r_{ab}{}^a
                                 \left[
                                  \frac{4N^{\frac12}}{q^{\frac12}} J^b
                                  - q^{cd} \p_c \p_dN^{b}
                                  - \eps^3 N \p^b r^{de}{}_c q_{de} N^c
                      \right. \right.
                                                             \nonumber \\
                &   & \left.
                      - q^{-\frac12} \p_b p^{de}q_{de}
                                 \right]
                      \left.
                      - \frac1{\eps} r_{ab}{}^c q^{ab}
                        \p_c (2N\dot U + \p_d N^d)
                         \right\}
                                                 \;\; d\Sigma
                                                              \nonumber \\
                &   & + \;\; \eps \;\; regular \;\; algebraic
                        \;\; terms \;\; in \;\; the \;\; fields
                        \;\; and \;\; in
                                                               \nonumber \\
                &   &   \;\;\; first \;\; derivatives \;\; of
                        \;\; U \;\; and \;\; N^a.
\eea

We observe that there are  two different types of singular terms, one
is a factor of $r^{ab}{}_a$ the other of $r^{ab}{}_c q_{ab}$. To
eliminate the first term we choose a gauge (i.e. a selection of lapse
and shift) that makes $r^{ab}{}_a \equiv 0$.  But to achieve this we
need $U$ to satisfy:

\be
     \Delta U = \rho + \eps^2 F(\eps, r^{ab}{}_c, p^{ab}, \p_c U),
\ee

since otherwise the constraint equation (16) would imply that $\p_b
r^{ab}{}_a$ is different form zero.  We then adjust $N^a$ such that
$\dot r^{ab}{}_b = 0$.  This implies the following equation for $N^a$:

\begin{equation}
\partial_c D^c N^b - \partial^b D_c N^c -
4(J^b + \partial^b N\dot{U}) +2\eps^2Nq^{cb}N^d\partial_c\partial_dU -
\eps^2 G^b =0.
\end{equation}

A necessary condition for this equation to have solutions is that the
above equation for $U$ be satisfied. In fact, from Bianchi identities
it follows that the divergence of the above equation vanishes
identically, regardless of $N^a$.  Equation (25) does not fix $N^a$
completely; in fact as it stands it is not even elliptic. We use the
remaining freedom in the lapse to choose

\be
        D_d N^d = -2 N \dot U,
\ee

getting rid of the second term.  To see that this is possible we
add to equation (25) the term $q^{bc} \p_c(4N\dot U + 2 D_dN^d)$, thus
eliminating the terms with $\dot U$ from it and also rendering it
elliptic:

\be
       \p_c D^c N^b + \p^b D_c N^c
                           - 4 J^b
 + 2\eps^2 N q^{cb}N^d \p_c \p_d U  = \eps^2 G^b
                                                (\eps, r^{ab}{}_c,
                                                 p^{ab}, \p_c U).
\ee

Since the divergence of equation (25) vanishes identically the
divergence of equation (27) must then be:

\be
         0 \equiv \Delta (4 N \dot U + 2 D_d N^d).
\ee

Uniqueness of solutions to Laplace's equation then implies equation
(26).

We have thus eliminated from the time derivative of the energy all
$\eps$-singular terms. The price we pay for this regularization is that
now we must consider a mixed symmetric-hyperbolic-elliptic system of
equations, the hyperbolic part being equations (4, 14, 15), the
elliptic part being equations (24,27) \footnote{Equations (24,27) form
an elliptic system for $(U,N^a)$ as long as the tensor $q^{ab}- \eps^4
N^a N^b$ is positive definite. This in turn is a condition (for $\eps
\not = 0$) on how big $J^a$ is.}.  The initial constraint~\footnote{ We
call constraint equations those which only involve initial data, and
which once they are solved at the initial slice they remain valid for
all later times by virtue of the evolution equations.} equations are
now (17,18) and $r^{ab}{}_a = 0$. The time derivative of $U$ can be
eliminated from all the equations by using equation (26).  The authors
do not know of any general treatment of mixed
symmetric-hyperbolic--elliptic systems, and so in what follows we
present an argument leading to establish an energy estimate, and thus
existence of solutions for mixed systems. Because the scope of this
argument is broader than the topic here treated, and it involves
further techniques, we shall give the details elsewhere.

The argument is as follows: If we had G\aa rding estimate for the
elliptic variables in term of the hyperbolic ones, following the
procedure in [1] with only slight modifications we would arrive to an
energy estimate for just the hyperbolic variables --in a sense the
elliptic variables are just non-local (but smooth) functionals of the
hyperbolic ones--, and the existence of solutions would follow in the
usual way.  Thus we just need G\aa rding's estimate for the elliptic
fields, and for that we need to make sure the elliptic equations are
injective at each time slice.  We consider, for given initial data for
the hyperbolic fields, the elliptic system as depending on two
parameter families, $\eps$, and $t$, the evolution time.  It is easy to
see that for $\eps = 0$ the elliptic system --with suitable boundary
conditions-- has a unique solution, regardless of what the hyperbolic
fields are. Since the space of injective operators is open, then given
any initial data there will exist an $\eps_0$ such that for all $\eps
\in \lbrack 0, \eps_0 \rbrack$ the elliptic part of the system is
injective at $t=0$.  But for the same property, and since we are
assuming smoothness on $t$ (through the a priori smooth dependence of
the hyperbolic fields on $t$), there will exist a $T_0 > 0$ such that
the elliptic part of the system will remain injective for all times
smaller than or equal to $T_0$.  Thus we do have G\aa rding's estimate
during a finite period of evolution.

We have established an a priori energy estimate for general relativity
which remains valid even at the $\eps = 0$ limit. We remark that the
matrix $A^0_{ij}$ that appears in our equations is regular in $\eps$
times those dynamical variables whose equations are $\eps$-singular,
and so as remarked in [1], it is possible to get estimates which only
involve spatial derivatives of the fields in the energy
norms~\footnote{That is, in this case it is possible,  by using the
evolution equations, to trade time derivatives by spatial derivatives
of the variables without getting $\eps$-singular contributions.}.
Therefore the initialization process that we carry out bellow is not
needed as far as to get a priori estimates for norms involving only
spatial derivatives.  Nevertheless, some initialization  is needed to
obtain regularity in time of solutions, for in the singular equation
case one can not use the equations to conclude smoothness in time from
smoothness in space.

%%%%%%%%%%%%%%%%%%%%%%%%%%%%%%%%%%

\section{Initialization}

For the reason given above, it is important to control the time
derivative of the dynamical fields, that is, to get a priori estimates
for energy expressions which include $L^2$ norms of these time
derivatives.  This gives boundedness in time of the family of
solutions.

We now study the conditions on the initial data for these energy norms
to be initially bounded; that is all what is needed as made clear by
the results obtained in the last section. Obviously the terms having no
time derivatives of $p^{ab}$ and $r^{ab}{}_c$ do not have any singular
dependence on $\eps$. We first encounter problems with the terms $\dot
p^{ab} p_{ab} + \dot r^{ab}{}_c r_{ab}{}^c$, and therefore we must
choose the initial data --to initialize it-- so that they are regular.
Writing the initial data as, $U|_0 = U_0 + \eps U_1 +
...$, etc. we have that the condition for $\dot r^{ab}{}_c$ to be
regular at the initial time is:

\be
    \p_c p_0^{ab} = - e^{d(b} \p_c \p_d N_0^{a)},
\ee

that is,

\be
     p_0^{ab} = - e^{d(b} \partial_d N_0^{a)}.
\ee

Notice that this condition is consistent, since using the ~zeroth order
part of equation (27), we get:

\be
        \p_a p_0^{ab} = -\frac12 e^{db} \p_d \p_a N_0^a
                        -\frac12 e^{da} \p_d \p_a N_0^b
                      = - 2 J^b,
\ee

that is the ~zeroth order term of the vector constraint, equation (17).

The condition for $\dot p^{ab}$ to be regular is that

\be
     e^{cd}\p_d r_0^{ab}{}_c =0,
\ee

which implies that initially we must take $h_0^{ab} \equiv 0$.

If further smoothness in time is required for some application, then
one has to consider further time derivatives in the energy norm, and continue
the initialization procedure.

In the compact hypersurface case
further orders give rise to equations that can be solved at each order
without apparent obstructions, so in that case one would conjecture
infinite smoothness in time.  Existence of solutions to this finite
hierarchy of equations and of the constraints for the asymptotically
flat case deserves further study.
See references [6,7] for  different approaches
to get these conditions, and the existence of solutions to them.

\section{Conclusion}

We have shown the existence of an a priori energy
estimate~\footnote{Notice that our final (regular) system is a mixed
one: symmetric hyperbolic for the dynamical fields variables, elliptic
for the lapse and shift fields. In this case to establish  the a priori
energy estimate one needs G\aa rding's a priori estimate for the
elliptic part.} which holds for any value of the parameter $\eps$,
including the zero.  These estimates are an important step towards the
proofs of  existence\footnote{To be precise, besides the a priori
estimate to establish existence one needs an approximating sequence of
trial functions.} and smoothness of slow solutions, i.e. solutions
satisfying the initialization conditions, for any given matter source
system which is itself symmetric hyperbolic, which is regular in the
gravitational variables we have used, and which has a non-relativistic
current.  In particular this is the case for perfect fluid sources with
any non-relativistic initial configuration.

Which ~setting can we embed our calculation in to obtain complete
results on the existence and $\eps$-smoothness of near Newtonian
solution to Einstein's equations?
It is clear that the main obstacle for that is to find a ~setting where
terms either make a negative or null contribution to the energy estimate,
as to justify the fact that we have ignored them.
The ~simplest case where this is so is when there is no boundary, as would
be a spacelike closed cosmology ~admitting a Newtonian cosmology limit.
This would also be the case if we had an initial-boundary value formulation
of general relativity ~admitting maximally dissipative boundary conditions.
{}~Unfortunately this problem, which is badly needed for making confident
numerical computations, is not solved yet, partially due to the presence
of constraints, and partially due to the fact that very little is known about
the initial-boundary problem for nonlinear hyperbolic systems.\footnote{One of
the authors has already some results for the linearized Einstein's
equations in the harmonic gauge.} For this case -we conjecture- one would
find uniformly locally smooth one parameter families of solutions for
any given (small enough) Newtonian solution.
Another way to state this
conjecture would be the following: {\sl Given any Newtonian solution and any
integer $m$, there exists a general relativistic solution which stays
near~\footnote{Near in the sense of the energy norm.} the Newtonian
solution to order $\eps^m$ for a finite time interval.}
The thir case of interest is the asymptotically flat one where again we
do not have to worry about boundary conditions. The main difficulty
here is that one can not use the impressive machinery of Weighted Sobolev
Spaces and their corresponding Sobolev inequality for unbounded
domains, for the use of radial functions as weights would introduce
{}~unavoidable
$\eps$-singular terms on the corresponding energy estimate. Thus, since the
Sobolev inequality is badly needed to handle non-linearities, we have
to resort to Sobolev spaces which also include some given number of time
derivatives and use the boudedness in time of the evolution region to
get the desired inequality. But this implies that we need initially some
smoothness in the time direction for our Sobolev norms to start finite,
that is we need a grater degree of initialization. Thus, it seems to be
the case that $\eps$-smoothness in the asymptotically flat case is tied
to the absence, or presence, of obstructions to solve the hierarchy of
initialization equations to the needed order.

One could raise the question of why we need to impose a gauge condition
to get this results, since after all the theory is gauge invariant. A
partial answer to this is that to establish the a priori estimates we
have obtained one has to treat the fields in some fixed gauge, since it
is easy to make diffeomorphisms which are singular with respect to
$\eps$ in the limit $\eps \rightarrow 0$. But it seems to us that still
our gauge conditions could be relaxed somehow, for instance it should
be enough to fix them up to some order in $\eps$ and not to all orders
as we have done~\footnote{ Note that we need to fix it to all orders to
ensure that our system is symmetric hyperbolic--elliptic.}.  We believe
that the only gauge (up to first order in $\eps$) in which the
equations are regular is the above one, although we do not have a proof
of that. Of course one can pretend less, that is smoothness of $\eps
p^{ab}$, and of $\eps r^{ab}{}_c$, in that case the only non-singular
term in equation (22) is the one proportional to $(\Delta U - \rho)$,
and so the gauge: $\Delta U = \rho$, $N^a = 0$ ~suffices, but this is not
optimal.

There are other issues which deserve further study.

For analytical and numerical studies of the characteristic problem (the
initial time formulation along a future light-cone), it is of interest
to treat the related Newtonian limit, see for instance [10], and
establish similar results to the ones here obtained. In particular this
should be important as another justification to pick initialized data
as data with little extra radiation apart from the one coming from the
matter sources.  For related results see [3].

Once one has control of the incoming radiation not generated by the
sources one can start to consider in a rigorous way the back reaction
of radiation on the sources. There are methods available to treat those
effects along the lines we have considered here, and we think they
deserve some attention.
%%%%%%%%%%%%%%%%%%%%%%%%%%%%%%%%%%%%%%%%%%%%%%%%
%\newpage
\appendix
\section{Dimensions}

%%%%%%%%%%%%%%%%%%%%%%%%%%%%%%%%%%%%%%%%%%%%%%%%

In this appendix we briefly introduce the concept of dimension or of
units for geometrical objects, and give the rules we followed for
assigning dimensions to several geometrical objects that we used in the
third section of the paper in order to be able to determine where
$\eps$ appears on the equations.  Let $M$ be a manifold, and $C(M,R)$
the algebra of smooth real functions on it. We introduce dimensions by
enlarging this algebra to the cartesian product of $C(M,R)$ with a
discrete abelian group, $G$.  Each element of this new algebra is then
a pair $(f,T)$, where $f \in C(M,R)$ and $T \in G$ is the dimension of
$f$.  The product is the usual one: $(f,T) \times (g,L) = (fg,TL)$ and
the sum is only defined within pairs with the same group element:
$(f,T) + (g,T) = (f+g,T)$.  We shall, as usual, omit the pair and only
write $f$ for $(f,T)$, and we will denote the projection to the group
entry as $[(f,T)] = [f]=T$.

We shall require the coordinate functions to have a dimension different
than unity, say $L$, and define the dimension of a vector $n^a$ to be
the element $[n^a]$ of $G$ such that when the vector acts on any
function, $f \in C(M,R) \times G$, it gives a function of dimension
$[n(f)] = [n^a] \frac1L [f]$.  This definition is equivalent to
assigning the same dimension to the components of the vector. Note also
that the Lie bracket of two vector fields yields a vector of dimension
equal to the product of their respective dimensions divided by the
coordinate functions dimension.

We define the dimension of a covector $m_a$ to be the element $[m_a]$
of $G$ such that when the covector acts on any vector, $n^a$ it gives a
function of dimension $[m_a n^a]$. We extend these definitions to
tensor fields in the obvious way.  Note that then the connection has
dimension of $L^{-1}$ and the Riemann tensor has dimension of
$L^{-2}$.  If a metric tensor is present, then to be consistent with
the formulas in a coordinate system, the metric --and therefore its
components-- has to have as dimension the group identity, $[g_{ab}] =
E$.  With this convention then the length of a vector has the same
dimension as the vector, and raising and lowering indices do not change
the dimension of the objects.

For our application we ~assign: to the function $t$ that ~foliates the
space-time the dimension $[t] = T$, to the unit normal to the surface,
$[n^a] = E$, to the time flow vector $[t^a] = \frac LT$, (so that $[t^a
\nabla_a t ] = E$), to the inverse of the speed of light, $[\eps] =
\frac TL$. The dimension of any other object is defined following the
rules stated above, in particular; $[q^{ab} \equiv g^{ab} - n^a n^b ] =
E$, since $t^a \equiv \frac {\bar N}{\eps} n^a + \bar{N}^a$, $[\bar N]
= E$, and $[\bar N^a] = \frac LT$, $[\bar K_{ab} \equiv 2 \bar
\nabla_{(a} n_{b)}] = \frac 1L$, $[\bar \pi^{ab} \equiv \frac
{\sqrt{\bar q}}{\eps} (\bar q^{ab} \bar K^c{}_c - \bar K^{ab})] = \frac
1T$.

\vspace{1cm}

%%%%%%%%%%%%%%%%%%%%%%%%%%%%%%%%%%%%%%%%%%%%%%%%

\noindent{\large \bf Acknowledgements}

%%%%%%%%%%%%%%%%%%%%%%%%%%%%%%%%%%%%%%%%%%%%%%%%

The authors thank U. Brauer, R. Geroch, V. Hamity, L. Reyna, and J.
Winicour for discussions and advice.

%\section{References}
\vspace{1cm}
{\Large \bf References}

\begin{itemize}

\item[$1)$] G. Browning, and H.O. Kreiss, SIAM J. Appl. Math., {\bf 42},
 704, 1982.

\item[$2)$] Y. Choquet-Bruhat, and T. Ruggeri, Commun. Math. Phys. {\bf
89}, 269, 1983.

\item[$3)$] V.H. Hamity, O.A. Reula, and J. Winicour, Gen. Rel. Grav.
{\bf 22}, 1, (1990).

\item[$4)$] H.O. Kreiss, SIAM J. Numer. Anal., {\bf 16}, 980, 1979.

\item[$5)$] H.P. K\"unzle, and J.M. Nester, J. Math. Phys.
            {\bf 25},  1009, 1984.

\item[$6)$] M. Lottermoser, {\sl \"Uber den Newtonschen Grenzwert der
Allgemeinen Relativit\"atstheorie und die Relativistische Erweiterung
Newtonscher Anfangsdaten}, Ph.D. thesis,
Ludwig--Maximilians--Universit\"at
M\"unchen, 1988, {\sl The post-Newtonian approximation for the
constraint
equations in General Relativity}, Max--Planck--Institut f\"ur
Astrophysik
preprint, 1990.

\item[$7)$] O.A.  Reula, Comm. in Math. Phys. {\bf 117}, 1988.

\item[$8)$] E. Tadmor, Comm. on P. and Appl. Math. {\bf 35}, 839,
1982.

\item[$9)$] R. Wald, {\sl General Relativity}, Univ. of Chicago Press,
1984.

\item[$10)$] J. Winicour, J. Math. Phys. {\bf 24},  1193, 1983,
            J. Math. Phys. {\bf 25},  2506, 1984,
            Gen. Rel. Grav. {\bf 19},  281, 1987, and
            J. Math. Phys. {\bf 28},  668, 1987.

\end{itemize}

%%%%%%%%%%%

\end{document}